\RequirePackage{etoolbox}
\csdef{input@path}{%
 {sty/}% cls, sty files
 {img/}% eps files
}%
\csgdef{bibdir}{bib/}% bst, bib files

\documentclass{imsart}
\pubyear{0000}
\volume{00}
\issue{0}
\doi{0000}
\firstpage{1}
\lastpage{1}

\usepackage{amsthm}
\usepackage{amsmath}
\usepackage{natbib}
\usepackage[colorlinks,citecolor=blue,urlcolor=blue,filecolor=blue,backref=page]{hyperref}
\usepackage{graphicx}
\usepackage{graphicx} % For graphics.
\usepackage{amsmath}
\usepackage{amssymb}
\usepackage{tikz}

\usetikzlibrary{arrows}

\startlocaldefs
\numberwithin{equation}{section}
\theoremstyle{plain}

\providecommand{\s}{{\bf s}} % For location s
\providecommand{\h}{{\bf h}} % For separation vector h
\providecommand{\f}{{\bf f}} % For spectral matrix f
\providecommand{\Z}{{\bf Z}} % For observations Z
\providecommand{\om}{{\boldsymbol\omega}} % For frequency omega
\endlocaldefs

\begin{document}

\begin{frontmatter}
\title{Bayesian Spectral Modeling of Microscale Spatial Distributions in a Multivariate Soil Matrix}
\runtitle{Bayesian Spectral Modeling of Microscale Spatial Distributions}
%\thankstext{T1}{Footnote to the title with the ``thankstext'' command.}

\begin{aug}
\author{\fnms{Maria A.} \snm{Terres} %\thanksref{t1}
\ead[label=e1]{materres@ncsu.edu}},
\author{\fnms{Montserrat} \snm{Fuentes} %\thanksref{t1}
\ead[label=e2]{fuentes@ncsu.edu}},
\author{\fnms{Dean} \snm{Hesterberg} %\thanksref{t2}
\ead[label=e3]{dean\textunderscore hesterberg@ncsu.edu}}
\and
\author{\fnms{Matthew} \snm{Polizzotto} %\thanksref{t2}
\ead[label=e4]{matt\textunderscore polizzotto@ncsu.edu}}

%\thankstext{t1}{Some comment}
%\thankstext{t2}{First supporter of the project}
%\thankstext{t3}{Second supporter of the project}
\runauthor{M. A. Terres et al.}

\affiliation{North Carolina State University, Department of Statistics\thanksmark{t1} and North Carolina State University, Department of Soil Science\thanksmark{t2}}

\address{Department of Statistics, North Carolina State University,
Raleigh, NC 27695-8203, \printead{e1}, \printead*{e2}}

\address{Soil Science Department, North Carolina State University,
Raleigh, NC 27695-7619, \printead{e3}, \printead*{e4}}
\end{aug}

\begin{abstract}
Recent technological advances have enabled researchers in a variety of fields to collect accurately geocoded data for several variables simultaneously. In many cases it may be most appropriate to jointly model these multivariate spatial processes without constraints on their conditional relationships. When data have been collected on a regular lattice, the multivariate conditionally autoregressive (MCAR) models are a common choice. However, inference from these MCAR models relies heavily on the pre-specified neighborhood structure and often assumes a separable covariance structure. Here, we present a multivariate spatial model using a spectral analysis approach that enables inference on the conditional relationships between the variables that does not rely on a pre-specified neighborhood structure, is non-separable, and is computationally efficient. Covariance and cross-covariance functions are defined in the spectral domain to obtain computational efficiency. Posterior inference on the correlation matrix allows for quantification of the conditional dependencies. The approach is illustrated for the toxic element arsenic and four other soil elements whose relative concentrations were measured on a spatial lattice. Understanding conditional relationships between arsenic and other soil elements provides insights for mitigating poisoning in southern Asia and elsewhere.
\end{abstract}

%\begin{keyword}[class=MSC]
%\kwd[Primary ]{60K35}
%\kwd{60K35}
%\kwd[; secondary ]{60K35}
%\end{keyword}


\end{frontmatter}
% % % % % % % % % % % % % % % % %
% % % % % % % % % % % % % % % % % % % % % % %
% % % % % % % % % % % % % % % % % % % % % % %
% % % % % % % % % % % % % % % % % % % % % % %
\section{Introduction}
Expansive spatial datasets have become more and more common as data have become easier to collect with improved technologies. In turn, this has created a need for computationally efficient modeling approaches that can accommodate these large datasets. Examples of such approaches include the predictive process \cite[]{banerjee2008gaussian}, nearest-neighbor Gaussian processes \cite[]{datta2014hierarchical}, partitioning of the spatial region \cite[]{kim2005analyzing}, covariance tapering \cite[]{sang2012full},  
and others. Each of these approaches exhibits unique strengths and weaknesses, as discussed by \cite{stein2014limitations}. In addition, for many modern datasets there may be interest in modeling multiple spatial variables jointly, treating them as dependent spatial random variables, in lieu of a regression approach where several variables are simply being conditioned on. However, few of the aforementioned methods can easily accommodate these multivariate spatial datasets. Development of computationally efficient approaches to handle such data has the potential to greatly improve the extent to which researchers can learn about the relationships between spatial processes.

When data have been collected on a spatial lattice, common in fields such as medical imaging and environmental science, some of the most common modeling approaches come from the family of conditionally autoregressive (CAR) models, laid out by \cite{besag1974spatial} and extended to the multivariate case (MCAR) by \cite{mardia1988multi}. At their heart, CAR models have been defined such that the spatial observation at any location will be normally distributed with a mean that is a weighted average of the neighboring observations. Defined through a series of conditional distributions, CAR models assume conditional independence between observations that are not spatially adjacent, and as such are special cases of Markov Random Field (MRF) models.  
In the literature MCAR models 
have been criticized for possessing a poorly identified and overly elaborate dependence structure, and multiple reparameterizations have been proposed to improve propriety and model behavior \cite[]{gelfand2003proper, sain2007spatial, zhang2009smoothed, sain2011spatial}.

Within the Bayesian literature, the most common approaches to modeling multivariate lattice data come from the family of MCAR models. Although originally proposed by \cite{mardia1988multi}, the most prevalent form is the adaptation by \cite{gelfand2003proper} ensuring distributional propriety. In this form the covariance structure is separable and can be decomposed into a covariance matrix describing the (non-spatial) relationship between the variables and a second covariance matrix describing the spatial dependence shared across all variables. Although this is computationally efficient, it is quite restrictive in its description of the marginal spatial dependencies of the variables. Alternative formulations, such as \cite{jin2005generalized} and \cite{jin2007order}, allow for non-separable formulations but become more computationally expensive. Finally, all of these approaches rely on a pre-defined neighborhood structure that limits the shape and extent of the spatial dependence in a way that is avoided by working with the spectral approach we propose here.

Although the MCAR models allow for consideration of multivariate spatial processes, they only maintain their computational efficiency if one is willing to assume a separable covariance structure. In addition, the Markovian structure of the model constrains spatial dependence to a set of neighbors pre-determined through an adjacency matrix. This is in contrast to geostatistical approaches where spatial dependence is assumed to decay as a smooth function of distance and the covariance parameters. 
In situations where there is interest in jointly modeling multivariate spatial lattice data while avoiding the separable Markovian structure and propriety issues inherent in MCAR models, spectral analysis procedures provide a natural framework to turn to. Computations are conducted after transforming the data into the ``spectral domain," allowing for greater efficiency as discussed in Section \ref{sec:spec}. The literature on spectral analysis techniques is prolific in time series contexts \cite[]{priestley1981spectral, koopmans1995spectral}, and also fairly common in the frequentist spatial literature \cite[]{stein1999interpolation}. However with a few exceptions, e.g. \cite{handcock1993bayesian}, \cite{reich2012nonparametric}, and \cite{stroud2014bayesian}, there has been relatively little work in this area when modeling spatial data within a Bayesian framework. 

In this paper we present a joint multivariate spatial model using spectral analysis procedures to gain computational efficiency. Each spatial variable has a spectral density controlling the marginal spatial covariance function, while a correlation matrix controls the relative cross-covariances between the variables. The model is fit in a Bayesian framework, allowing for uncertainty quantification in all aspects of the model. In particular, posterior examination of the correlation matrix provides inference on the nature of the conditional dependencies between the variables. 

This methodology is illustrated for an analysis of accumulation of potentially toxic arsenic in a soil matrix. Synchrotron X-ray fluorescence microprobe ($\mu$-XRF) analysis was used to map accumulated arsenic in relation to other chemical elements in thin coatings on a quartz sand grain collected from a soil sample. The technique produces multivariate spatial lattice maps that essentially reflect relative abundances of elements. Interest lies in understanding the conditional dependencies between arsenic and the other soil components, and whether these dependent components serve to mitigate or potentiate the accumulation of arsenic. 
Arsenic contamination of drinking water is a widespread human health concern, particularly in Asian countries where over 100 million people routinely consume dangerous levels of arsenic because of their reliance on arsenic-contaminated well water for drinking water \cite[]{ravenscroft2009arsenic}. 

Several methods of water treatment have been studied with varying success, including the introduction of additional chemical salts or solutions that will react with the arsenic \cite[]{jiang2012arsenic, komarek2013chemical}. However, because soils comprise multiple elements in multiple mineral and organic components, a more precise understanding of chemical reactions could be aided through a statistical description of the soil elements' dependencies. Unlike the common approach of modeling X-ray absorption spectra from soils to identify pure chemical species \cite[]{manceau2014estimating}, our approach aims to identify via element associations possible interactions between different soil components that produce uniquely complex species, or cause the reactivity of any pure soil species to differ from that of their model analogues studied in isolation of soil.

Studying the pairwise behavior of soil elements, a common approach for analysis of microscale soil chemical data, provides only a very limited view of their relationships, and neglects to account for interactions between the elements. This is in contrast to our hypothesis that interactions between soil components inferred from element pairs may depend on other co-localized element compounds. In order to capture these kinds of behaviors, it is necessary to have a model that is sufficiently flexible in its treatment of conditional dependencies. 

The model developed here is illustrated for arsenic and several co-localized elements. In natural systems, the mobility and toxicity of arsenic is largely controlled by various competing abiotic and biotic redox and adsorption processes involving organic matter and iron, aluminum, and manganese oxides \cite[]{borch2009biogeochemical}. Many of these reactions have been studied using single or binary mixtures of aqueous species, model minerals and/or organic components. In contrast, less mechanistic detail is known about arsenic behavior in soils, which comprise diverse assemblages of minerals and organic matter, and conditional dependency models would enable scientists to  better understand how arsenic behaves when multiple components are co-localized within complex soil matrices.

The lattice structure of the $\mu$-XRF data make spectral procedures a natural choice for model-fitting. This approach was previously explored by \cite{guinness2014multivariate} in a frequentist setup, but the methodology could not accommodate more than three spatial variables and lacked adequate measures of uncertainty. The need for additional model flexibility and improved uncertainty quantification indicated a fresh look at the problem from a Bayesian perspective. Unlike the model developed by \cite{guinness2014multivariate}, the modeling framework we outline can easily accommodate any arbitrary number of spatial variables. We illustrate the methodology with five soil elements, including the three that were previously analyzed, providing full descriptions of the uncertainty associated with our estimates and producing conditional dependence graphs that clearly illustrate the relationships between the variables. 

The remainder of this article is organized as follows. Section \ref{sec:spec} provides an overview of the spectral analysis approach in the spatial domain, outlining some properties of the commonly used approximations. In Section \ref{sec:model} the soil mineral data are presented as an illustrative example, and the modeling details are specified. In Section \ref{sec:results} the model results and potential implications are discussed. Finally, Section \ref{sec:concl} concludes with a brief discussion of the presented methodology and its potential for future work.

% % % % % % % % % % % % % % %
\section{Spectral Methods}
\label{sec:spec}
\subsection{Computing the Likelihood}
Consider a spatial process ${\bf Z}=(Z(\s_1), \hdots, Z(\s_N))'$ assumed to be a realization from a Gaussian process with stationary covariance function $c(\h)=\text{cov}(Z(\s), Z(\s+\h))$ that depends on parameters $\theta$. Let $\Sigma_\theta$ denote the corresponding covariance matrix. Then ${\bf Z}\sim N(\boldsymbol0, \Sigma_\theta)$ and the log-likelihood can be computed,
\begin{equation}
\log(p({\bf Z}|\theta)) = -\frac{1}{2}(\log\det \Sigma_\theta + {\bf Z}'\Sigma_\theta^{-1}{\bf Z})\label{eq:loglik}
\end{equation}  
where proportionality constants have been ignored. Normal likelihoods are generally easy to compute when $N$ is small and are commonly used in spatial modeling \cite[]{stein1999interpolation,cressie2011statistics, banerjee2014hierarchical}. However, the matrix inverse $\Sigma_\theta^{-1}$ requires $O(N^3)$ floating point operations (flops), rendering computation of this likelihood undesirable when working in large dimensions. When the observations are available on a lattice of size $N=n_1\times n_2$, denoted $\mathbb{J}_N$, then it is convenient to work in the spectral domain since the normal log-likelihood can be approximated using the Whittle likelihood \cite[]{whittle1954stationary, zimmerman1989computationally}. This approximation takes advantage of the computational efficiency of the Fast Fourier Transform (FFT), dramatically reducing computation time. This approach is outlined below.

Bochner's Theorem states that any stationary covariance function can be represented as an inverse Fourier transform,
\begin{equation}
c(\h) = \int_{\mathbb{R}^2} \exp(i\om'\h)dF(\om)\label{eq:spec.cov}
\end{equation}
where $\h$ is a separation vector for spatial locations $\s$ and $\s+\h$, and $\om=(\omega_1,\omega_2)$ is a bivariate spectral frequency. We assume there exists some continuous differentiable $f(\om)$ such that $dF(\om)=f(\om)d\om$, commonly referred to as the spectral density. The Spectral Representation Theorem states that the spatial process associated with this covariance function can similarly be represented,
\begin{equation}
Z(\s) = \int_{\mathbb{R}^2} \exp(i\om'\h)d\tilde{Z}(\om)\label{eq:spec.z}
\end{equation}
where $d\tilde{Z}(\om)$ have uncorrelated increments and $E|d\tilde{Z}(\om)|^2=f(\om)$. 

When the data have been observed on a grid there is inadequate information to fully recover the continuous process. In particular, if the data are observed at uniformly spaced locations with intervals of length $\delta$, i.e. the process can be written $Z(\delta{\bf x})$ for ${\bf x}\in \mathbb{Z}^2$, then the associated spectra are constrained to frequencies in the finite interval $-\pi/\delta\le \omega\le\pi/\delta$. This can be seen by observing that $\exp(i\om'\h)=\exp(i(\om+2\pi j/\delta)'\h)$, the so called ``aliasing" phenomenon. 

When working with lattice data the aliasing must be accounted for in the choice of the associated spectral density. One approach is to choose a spectral density defined on the real line, then accumulate the density over all aliased frequencies. An alternative is to work with a spectral density that has been explicitly defined to have support on $\om\in[-\pi/\delta, \pi/\delta]^2$. We follow the latter approach in our analyses, selecting the quasi-Mat\'ern spectral density introduced by \cite{guinness2014multivariate}. To ensure a real-valued process, the spectral densities must additionally be even functions symmetric around zero, which is again satisfied by the quasi-Mat\'ern spectral density. 
 
In practice we cannot compute the integrals in (\ref{eq:spec.cov}) and (\ref{eq:spec.z}), so we instead approximate them with discrete sums,
\begin{align}
c(\h) &= \sum\limits_{j\in\mathbb{J}_N} e^{i\om_j \h}f(\om_j)\label{eq:spec.z.approx}\\
Z(\s) &= \sum\limits_{j\in\mathbb{J}_N} e^{i\om_j \s}\tilde{Z}(\om_j)\label{eq:disc.period} %\\
%\om_j &= 2\pi j/T\nonumber
\end{align}
evaluated at the Fourier frequencies $\om_j =(2\pi j_1/n_1, 2\pi j_2/n_2)$, $j=(j_1,j_2)\in \mathbb{J}_N$. These approximations have a long history of use in spectral modeling of spatial processes, and their properties are well established \cite[]{whittle1954stationary, guyon1982parameter}.

Due to the lattice structure in the data, the corresponding covariance matrix $\Sigma_\theta$ will be block circulant and the FFT will effectively diagonalize $\Sigma_\theta$. This enables the log-likelihood to be rewritten in the spectral domain, with computation limited by the FFT requiring only $O(N\log N)$ flops, 
\begin{equation}
\log(p({\bf Z}|\theta)) = -\frac{1}{2}\left( \sum\limits_{j\in \mathbb{J}_N}\log(f(\om_j)) + \sum\limits_{j\in \mathbb{J}_N} F_N(Z_j)^*f(\om_j)^{-1}F_N(Z_j)\right)\label{eq:spec.lik}
\end{equation}
where $F_N({\bf Z})$ is an array denoting the 2-dimensional Fourier transformation of the lattice data $\bf Z$ such that the entry $F_N(Z_j)$ corresponds to the Fourier frequency $\om_j$. This equality can be shown in part by noting that the eigenvalues of $\Sigma_\theta$ correspond to the spectral density $f(\om)$ evaluated at each of the Fourier frequencies.  

The disadvantage of this approximation is that the discreteness of (\ref{eq:spec.z.approx}) and (\ref{eq:disc.period}) produces the generally undesirable property of periodicity over the range of the data. An adjustment is necessary to mitigate this feature. Some new approaches have recently been proposed in the literature involving an imputed expansion of the lattice such that the periodicity occurs beyond the domain of the observed data \cite[e.g.][]{guinness2014circulant, stroud2014bayesian}. However, we follow the more common approach of data tapering \cite[]{dahlhaus1983spectral, dahlhaus1987edge}. Intuitively, tapering dampens the data along the edges towards zero in a smooth way such that the lattice could be folded into a torus shape without introducing discontinuities. 

Specifically, we implement a cosine taper, or Tukey taper \cite[]{tukey1967introduction}, defined as,
\begin{align}
w_d(j) & = 
\begin{cases}
\frac{1}{2}(1+\cos(\frac{2\pi}{r}(j-\frac{r}{2}))), & 0\le j< \frac{r}{2}\\
1 ,& \frac{r}{2}\le j < 1-\frac{r}{2}\\
\frac{1}{2}(1+\cos(\frac{2\pi}{r}(j-1 + \frac{r}{2}))) , & 1- \frac{r}{2}\le j \le  1\\
\end{cases}
\end{align}
for dimensions $d=1,2$ and $j=0/n_d, \hdots, (n_d-1)/n_d$. The parameter $r$ controls the extent of the tapering, typically chosen to be 5-10\% of the observations at each boundary. The original data $Z_j=Z_{j_1,j_2}$ in the likelihood  in (\ref{eq:spec.lik}) is then replaced with the tapered data $Z_{j_1,j_2}\times w_1(j_1)\times w_2(j_2)$, and a multiplicative adjustment of $\prod\limits_{d=1}^2\sum\limits_{j=1}^{n_d}w_d(j)^2$ is incorporated into the $F_N(Z_j)$ terms.

When the data are multivariate, with lattice observations for $M$ spatial variables $\Z =(\Z^{(1)}, \hdots, \Z^{(M)})$, the multivariate process will require a positive definite $M\times M$ matrix of spectral densities, $\f(\om)$. The matrix entries along the diagonal, $\f_{m,m}(\om)$, dictate the marginal covariances for each of the variables. Similarly, the matrix entries on the off-diagonal, $\f_{m,m'}(\om)$, dictate the cross-covariances between each pair of variables. The likelihood can then be computed similarly to the univariate case,
\begin{equation}
\log(p({\bf Z}|\theta)) = -\frac{1}{2}\left(\sum\limits_{j\in\mathbb{J}_N}\log\det\f(\om_j) + \sum\limits_{j\in\mathbb{J}_N}F_N(\Z_j)^*\f(\om_j)^{-1}F_N(\Z_j)  \right)\label{eq:spec.lik.mvt}
\end{equation}
where $F_N(\Z^{(i)})$ is an array denoting the 2-dimensional Fourier transformation of the lattice data for spatial variable $\Z^{(i)}$ such that entry $F_N(\Z^{(i)}_j)$ corresponds to the Fourier frequency $\om_j$, the $M\times 1$ vector $F_N(\Z_j)=(F_N(\Z^{(1)}_{j}), \hdots, F_N(\Z^{(M)}_{j}))'$ concatenates the elements of these matrices corresponding to 
the frequency $\om_j$ for each spatial variable,
and $F_N(\Z_j)^*$ is the analogous vector corresponding to the complex conjugate transpose of the Fourier transformed lattice data. Data tapering can be conducted in the same manner as in the univariate case.

Similar to the univariate case, computation of the multivariate likelihood (\ref{eq:spec.lik.mvt}) is limited only by the computation of the FFT on each of the spatial variables and the inversion of the $M\times M$ matrix $\f(\om)$.
% % % % % % % % % % % % % % % % % % % %
\section{Multivariate Spatial Sand Grain Model}
\label{sec:model}
% %
% % % % % % % % % % % % % % %
\subsection{Sand Grain Data}
Data generated for this analysis were obtained by introducing solutions of arsenic to a soil sand grain and subsequently assessing the microscale spatial distributions of arsenic and a number of native soil elements. The sand grain analyzed was separated from a surface soil sample collected from a forest at the Central Crops Research Station in Clayton, NC. %The soil sample was dispersed by shaking in deionized water, and sand grains ($>50~\mu$m) were collected on a Teflon screen and dried under dinitrogen gas. One sand grain comprising an unreactive quartz core covered with a thin (10-20 $\mu$m) pedogenic coating of secondary minerals and organic matter was lodged between two walls of a polypropylene sample holder.  
The spatial distributions of elements %with atomic numbers between Ca and As 
were mapped using $\mu$-XRF %at 12 keV or at the arsenic(V) white line (11.875 keV) 
using Beamline X27A at the National Synchrotron Light Source (NSLS), Brookhaven National Laboratory. 
Our analysis focuses on element maps collected after sequential treatments of the grain with 100 and 1000 $\mu$M arsenic (III) treatments.
%Element mapping was repeated after sequential arsenic(III) treatments with 100 $\mu$M and 1000 $\mu$M KH$_2$AsO$_3$ solutions.  
A 350$\times$450 $\mu$m region of interest (ROI) was mapped using an X-ray beam of approximately 10$\times$10 $\mu$m$^2$ to yield a 35 $\times$ 45 pixel array. Resulting elemental maps reflect the relative abundance of elements within each 10$\times$10$\times\sim$15 $\mu$m$^3$ voxel analyzed by the incident X-ray beam.

We focus our analysis on the abundant soil matrix element arsenic (As), as well as four trace elements: iron (Fe), chromium (Cr), nickel (Ni) and zinc (Zn). The elements are modeled on the log-scale and, since our interest lies primarily in understanding the dependence structure between these elements, we center each spatial process by subtracting the sample mean and assume zero means in our modeling. Arsenic and iron were most abundant, averaging 7.23 and 9.47 on the log-scale, with the other elements averaging between 5.35 and 5.60. To specify the model, let ${\bf Z}^{(m)}(\s)$ be the centered log fluorescence signal at location $\s\in\mathbb{J}_N$ for soil element $m$, for $m=1,...,M$, with $M=5$. The ${\bf Z}^{(m)}(\s)$ are spatial quantities each assumed to be a realization from a Gaussian process and are modeled jointly in the spectral domain, as described in Section \ref{sec:spec}.

\begin{figure}[h!]
\centering
\includegraphics[width = 6.65in] {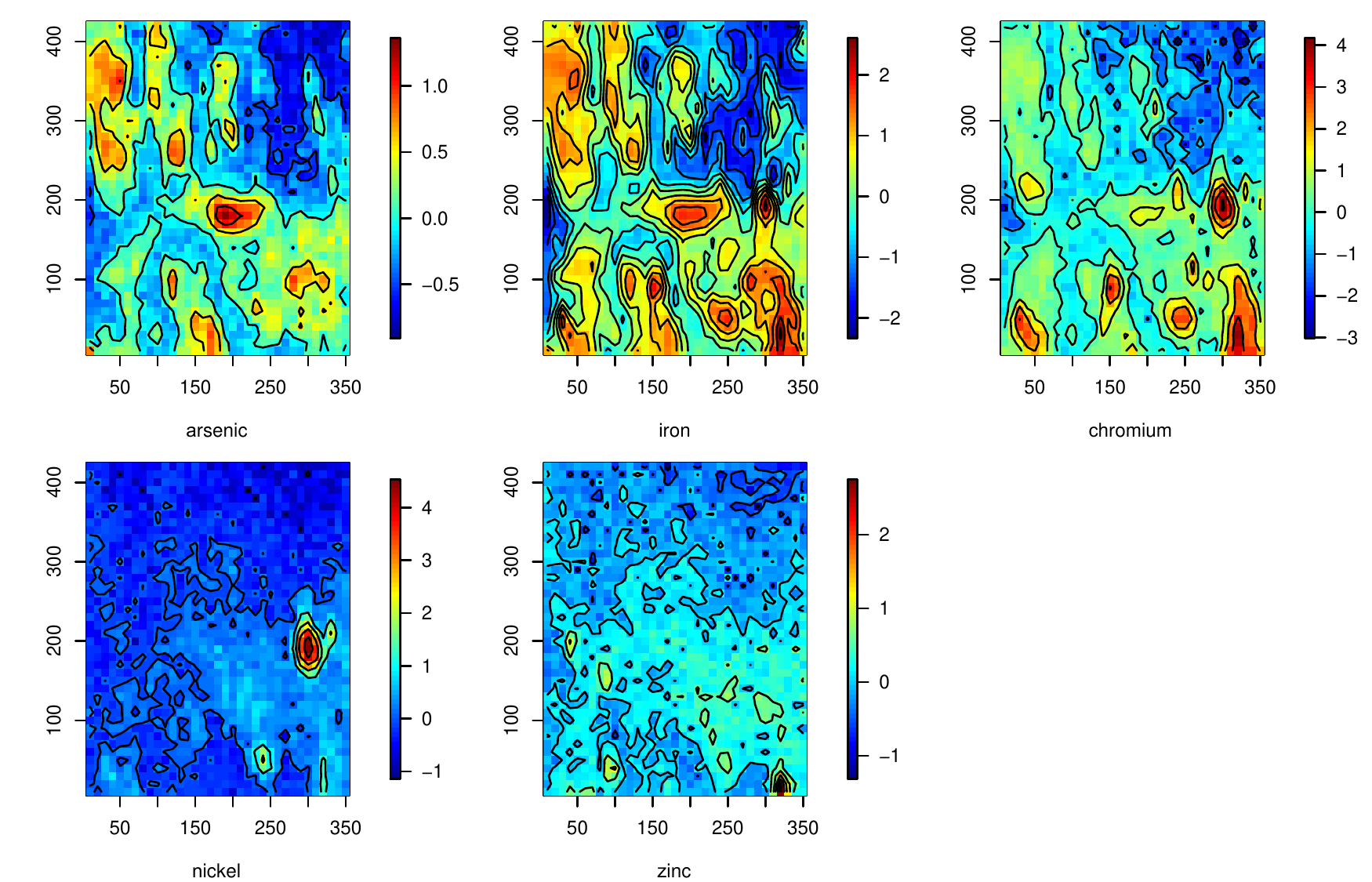}%{SoilData5_AsFeCrNiZn_bw_3}
\caption{The logged and centered soil components arsenic, iron, chromium, nickel, and zinc.}
\label{fig:data}
\end{figure}

The spatial maps of these five soil elements are provided in Figure \ref{fig:data}. Note that some concentrated hotspots of iron, chromium and nickel (e.g. at coordinates (300,200)) are known contaminants from stainless steel deposited on the sample during handling prior to reaction. The correlation between arsenic, iron, and to some extent chromium is readily apparent, with similar spatial features throughout the region. The correlations with nickel and zinc are less apparent. Through our joint modeling procedure we seek to make inference regarding the conditional dependencies exhibited by these five elements.

% % % % % % %
\subsection{Marginal Spectral Densities}
Recall that the covariance function, $c(\h)$, for each soil element is defined in the spectral domain using a spectral density, $f(\om)$.  The spectral densities are constrained to be even and to have support on $[-\pi/\delta,\pi/\delta]^2$ in order to produce real-valued realizations and to avoid the aliasing effect that occurs with lattice observations, as described earlier. With these requirements in mind, we focus on the non-separable quasi-Mat\'ern spectral density defined on a lattice,
\begin{equation}
f(\om)= f(\omega_1,\omega_2) = \frac{\sigma^2}{(1+(\alpha/\delta)^2(\sin^2(\delta\omega_1/2) + \sin^2(\delta\omega_2/2)))^{\nu+1}}\label{eq:specdens}
\end{equation} 
for $\om=(\omega_1,\omega_2)\in[-\pi/\delta, \pi/\delta]^2$ as described by \cite{guinness2014multivariate}. This spectral density is attractive in part because it approaches the spectral density of the isotropic Mat\'ern covariance function as $\delta\to 0$. The parameter $\alpha$ functions as a range parameter, indicating the distance at which spatial dependence becomes negligible. The parameter $\sigma^2$ functions as a variance parameter, controlling the magnitude of the spatial correlation. Due to concerns regarding lack of identifiability between the parameters \cite[]{zhang2004inconsistent}, we fix $\nu\equiv 1$ in our model fitting. We additionally assume that each $10\times10\mu m$ grid cell represents one unit of distance, setting $\delta\equiv1$.
% %
\subsection{Cross-Covariance Structure}
To guarantee a well defined spatial process the $M\times M$ matrix $\mathbf f(\om)$, corresponding to the cross-covariance matrix in the spatial domain, must be positive definite. To guarantee this, we follow the approach of \cite{guinness2014multivariate} and write 
\begin{align}
\mathbf f(\om) &= \textnormal{diag}(f^{1/2}(\om)) \boldsymbol\rho(\om)\textnormal{ diag}(f^{1/2}(\om))
\end{align}
where diag$(f^{1/2}(\om))$ is a diagonal matrix with diagonal entries $f^{1/2}_m(\om)$, the square root of the marginal spectral density for the $m$th element as defined in (\ref{eq:specdens}), for $m=1,\hdots,M$. Here, $\boldsymbol\rho(\om)$ is the coherence matrix, a correlation matrix with ones on the diagonal and elements $\rho_{m,m'}(\om)$ describing the correlation between components $m$ and $m'$. This is analogous to decomposing a covariance matrix into a correlation matrix and vectors of standard deviations, where here the standard deviations are the square roots of the spectral densities instead of the square roots of the variances. 

Even in relatively low dimensions, modeling the matrix $\boldsymbol\rho(\om)$ will require the estimation of many correlation functions $\rho_{m,m'}(\om)$. For example, in our example with 5 soil elements there would be $5\times(5-1)/2=10$ functions that need to be estimated in the $\boldsymbol\rho(\om)$ matrix. Estimating this many functions may rapidly become prohibitive as the dimension increases. As an alternative, we propose 
\begin{align}
\mathbf f(\om) &= \textnormal{diag}(f^{1/2}(\om)) \boldsymbol\rho\textnormal{ diag}(f^{1/2}(\om))\label{eq:constrho}
\end{align}
where the matrix $\boldsymbol\rho$ is constant across frequencies. This assumption implies that for any pair of soil elements, the spatial dependence described by the cross-covariance functions will be dictated solely by the pair of marginal covariance functions and a multiplicative factor $\rho_{m,m'}$. While this assumption is primarily motivated by the need for computational efficiency, the resulting model is also better identified and still defines a very flexible framework deemed sufficient to describe the $\mu$-XRF data.  

Decomposing $\f(\om)$ as in (\ref{eq:constrho}) additionally simplifies the computation of the likelihood. Specifically, it is straight forward to show that $\det\f(\om)=(\det\boldsymbol\rho)\prod\limits_{m=1}^Mf_m(\om)$ and to rewrite the inverted matrix $\f(\om)^{-1}=\textnormal{diag}(f^{-1/2}(\om))\boldsymbol\rho^{-1}\textnormal{diag}(f^{-1/2}(\om))$. Then, substituting into the multivariate likelihood from (\ref{eq:spec.lik.mvt}) we have,
 \begin{align}
  \log(p({\bf Z}|\theta)) = &-\frac{1}{2}\Big( N\log\det\boldsymbol\rho + \sum\limits_{j\in\mathbb{J}_N}\sum\limits_{m=1}^M \log f_m(\om_j) \\&+ \sum\limits_{j\in\mathbb{J}_N} F_N(\Z_j)^* \textnormal{diag}(f^{-1/2}(\om))\boldsymbol\rho^{-1}\textnormal{diag}(f^{-1/2}(\om)) F_N(\Z_j)\Big)\nonumber
 \end{align}
 Note, during model fitting the inversion and determinant of $\boldsymbol\rho$ need only be computed once at each iteration since it does not depend on the frequency. 

\subsection{Conditional Independencies and the Correlation Matrix}
\label{subsec:condl}
As described by \cite{guinness2014multivariate}, conditional independencies between the soil components can be explored through examination of $\mathbf f(\om)^{-1}$, or equivalently through examination of $\boldsymbol\rho^{-1}$ under our proposed parameterization. The Bayesian paradigm allows us to consider uncertainties for each element of the correlation matrix enabling examination of the implied conditional independencies, similar to the approach followed by \cite{hoff2007extending}. Specifically, consider a vector of normal random variables $(z_1,z_2,\hdots,z_n)'$ with associated correlation matrix ${\bf C}$. Then in the conditional distribution for $z_j$ given the other variables, $p(z_j|z_{-j})$, the ``coefficients" on $z_{-j}$ will be ${\bf C}_{[j,-j]}{\bf C}^{-1}_{[-j,-j]}$. The sign of these ``coefficients" and whether the credible intervals overlap zero will provide inference on the graph of conditional dependence between variables. In this manner, the conditional independencies between the soil elements will be explored in our model through the correlation matrix $\boldsymbol\rho$. 
% % % % % % %
\subsection{Prior Specification and Model Fitting}
The following prior distributions are assumed for the parameters,
\begin{align*}
\alpha_m|s^2 & \sim TN_{(0,\infty)}(0, s^2)\\
s^2 & \sim IG(2, 2)\\
\sigma_m^2|\nu_0,\sigma_0^2 & \sim IG(\nu_0/2, \nu_0\sigma_0^2/2)\\
\sigma_0^2|\nu_0 & \sim IG(2,2)\\
p(\nu_0) & \propto e^{-\nu_0}\\
p(\boldsymbol\rho) & \propto 1, ~~\boldsymbol\rho\in\mathcal{R}^5
\end{align*}
where $\nu_0\in \mathbb{Z}^+$, and $\mathcal{R}^5$ is the space of all $5\times 5$ positive definite correlation matrices. Each spatial process $Z_m$ has a unique pair of parameters $(\alpha_m, \sigma^2_m)$, but hyperpriors are placed on these parameters to facilitate sharing of information across soil elements. The prior on $\boldsymbol\rho$ follows the suggestion of \cite{barnard2000modeling}, who similarly decomposed a covariance matrix into standard deviations and a correlation matrix.

We make inference on the model parameters through Markov chain Monte Carlo (MCMC). The $\alpha_m$ parameters were updated with a random walk Metropolis Hastings step with the proposal variance tuned during a burn-in period to achieve acceptance rates between 0.3 and 0.5. The $\sigma_m^2$ parameters are also updated with a Metropolis Hastings step, but in this case a more clever proposal distribution can be used. If the spatial process $Z_m$ were being modeled marginally, then $\sigma_m^2$ would be conjugate,
\begin{equation*}
\sigma^2_m|\theta \sim IG(\frac{\nu_0 + J}{2}, \frac{\nu_0\sigma_0^2 + \sum\limits_{j\in\mathbb{J}_N}F_N(\Z_j^{(m)})^*F_N(\Z_j^{(m)})(1+\alpha_m^2(\sin^2(\omega_{1j}/2) + \sin^2(\omega_{2j}/2)))^2}{2})
\end{equation*}
where $\theta$ is the collection of all other parameters. This conjugacy breaks down in the multivariate case, but we utilize this marginal conjugate distribution as a proposal in the Metropolis Hastings step with good results. The $\boldsymbol\rho$ matrix is updated with a griddy Gibbs step \cite[]{ritter1992facilitating}, as suggested by \cite{barnard2000modeling}. The hyperparameters $s^2, \nu_0, \sigma^2_0$ are updated via straightforward Gibbs steps.

% % % % % % % % % % % % % % % % % % % %
\section{Analysis Results}
\label{sec:results}
\begin{figure}[h!]
\centering
\includegraphics[width = 6in]{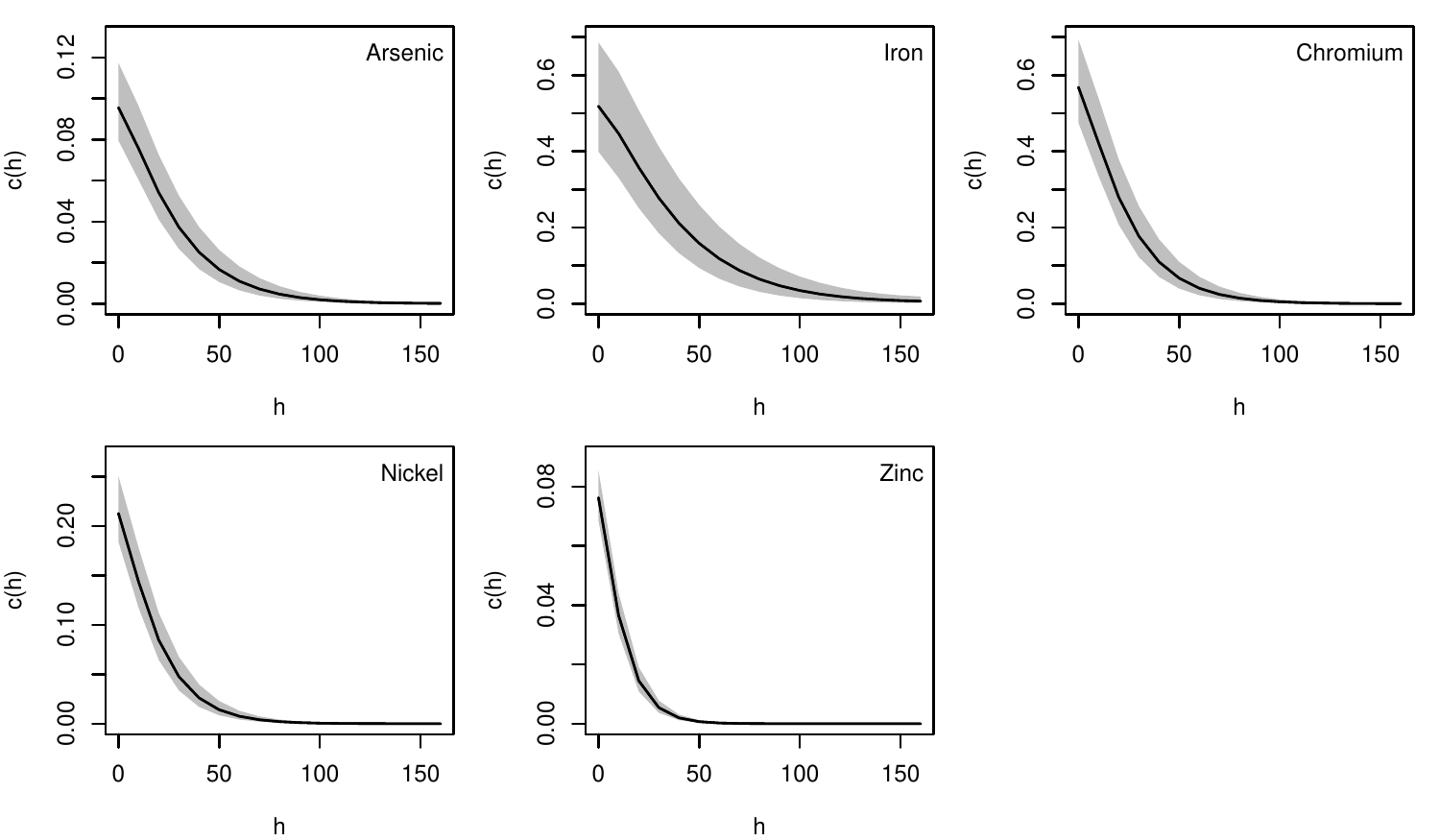}
\caption{Estimated covariance functions (95\% pointwise intervals in grey) for each of the five soil components, plotted for different distances $h$ ($\mu$m).}
\label{fig:covar}
\end{figure}

The model described above was fitted to the $\mu$XRF data with a 10\% taper yielding covariance and cross-covariance functions for each of the five soil elements and the ten soil element pairs. To check for sensitivity, the model was additionally run for a 5\% and 15\% taper with comparable results (not shown here). The covariance functions are provided in Figure \ref{fig:covar} and indicate varying marginal variances and spatial ranges across the soil elements, with iron having the longest spatial dependence and zinc having the shortest. The ability to capture this variability in spatial dependence is a direct outcome of the flexible spectral specification of the model and would be far more difficult to capture in a MRF model with a pre-specified neighborhood structure. In addition, the Bayesian approach provides straightforward descriptions of the uncertainty surrounding these covariance functions, shown here through grey shading representing 95\% pointwise credible intervals. 

\begin{figure}[h!]
\centering
\includegraphics[width = 7.5in, angle=90]{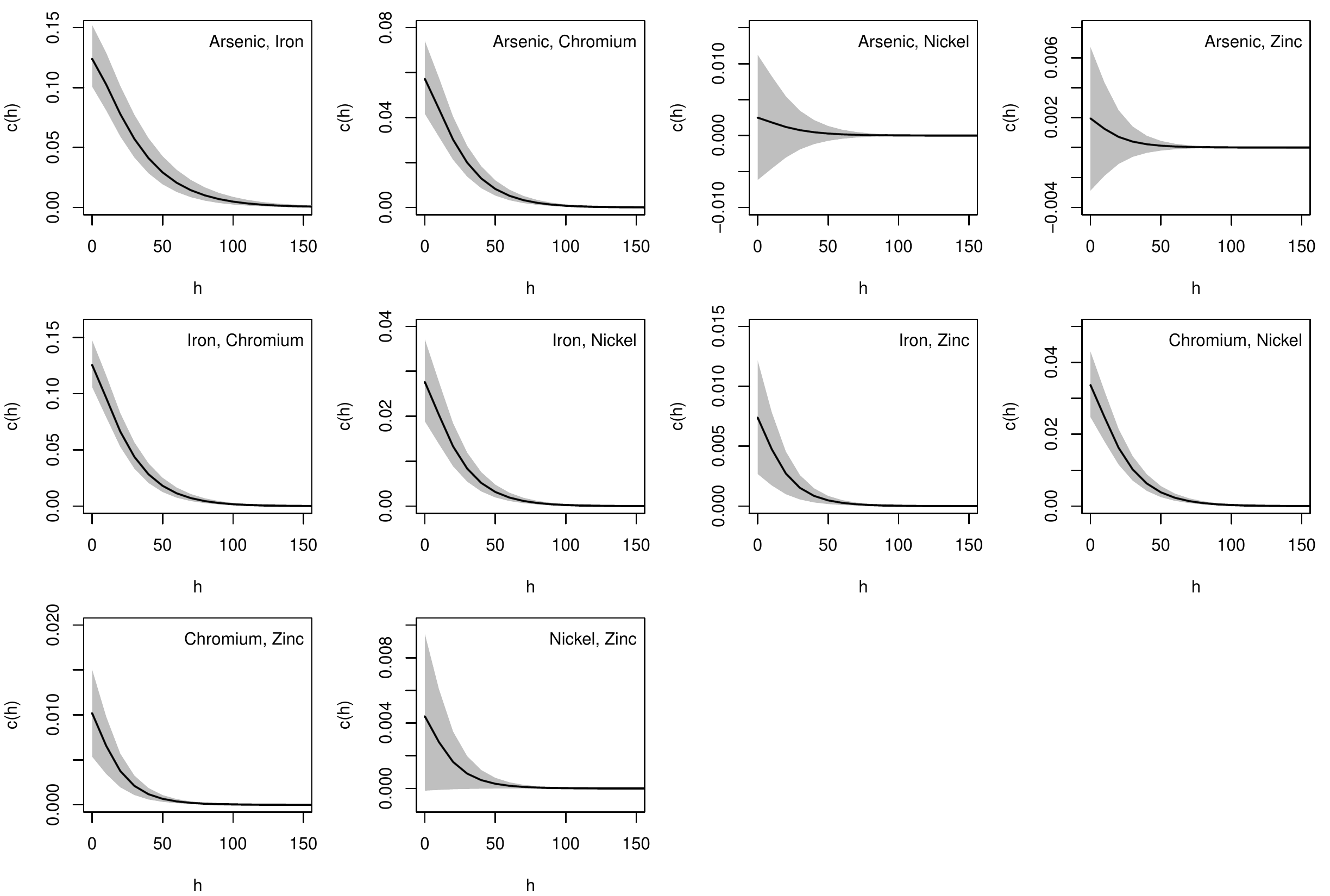}
\caption{Estimated cross-covariance functions (95\% pointwise intervals in grey) for each of the soil component pairs, plotted for different distances $h$ ($\mu$m).}
\label{fig:crosscovar}
\end{figure}

The cross-covariance functions are provided in Figure \ref{fig:crosscovar} and similarly illustrate the benefits of this spectral Bayesian modeling approach. In particular, the credible intervals for the cross-covariances between arsenic/nickel, arsenic/zinc and nickel/zinc clearly overlap zero. If one were to plot only the estimated covariance functions, then it would appear that these soil elements are minimally positively correlated with one another. However, the 95\% credible intervals enabled by the Bayesian framework make it quite clear that one cannot infer any pairwise relationships between these pairs of elements. 

% % % % %
% % % % %

\begin{figure}[h!]
\centering
\includegraphics[width = 7.5in, angle=90]{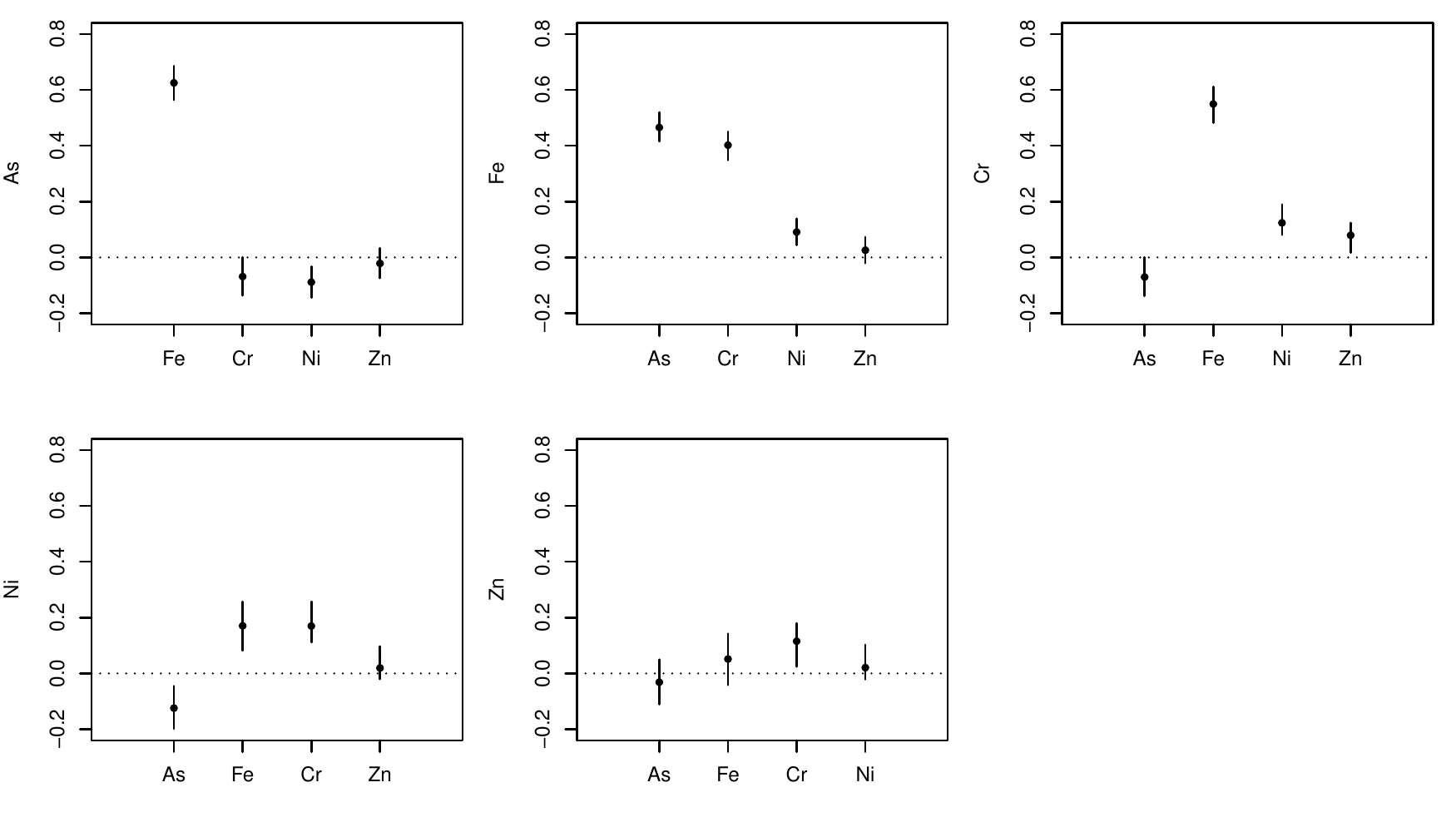}
\caption{Estimated ``coefficients" on $z_{-j}$ describing conditional dependencies between the soil components, defined to be $\Sigma_{[j,-j]}\Sigma^{-1}_{[-j,-j]}$.}
\label{fig:condl}
\end{figure}

In addition to learning about the pairwise relationships between these soil elements, researchers are particularly interested in learning multi-element effects on the chemical reactivity of an element, i.e., the nature of pairwise relationships when we condition on the presence of other elements in the sample. Using the approach outlined in Section \ref{subsec:condl}, for each soil element we consider the ``coefficients" that would be placed on the other soil elements in the conditional distribution implied by the fitted joint model. Posterior estimates and credible intervals for these ``coefficients" are provided in Figure \ref{fig:condl}, with each subplot corresponding to one of the five soil elements. For example, looking at the first subplot one can see that arsenic has a strong relationship with iron and a slight negative relationship with nickel, but does not have a significant relationship with chromium and zinc. In conjunction with Figure \ref{fig:crosscovar}, we can infer that arsenic and chromium will be positively correlated when examined pairwise, but that this relationship disappears when the other soil elements are accounted for.

% % % % %

\begin{figure}
\centering
\begin{tikzpicture}[>=stealth',shorten >=1pt,auto,node distance=3cm,
  thick,main node/.style={circle,draw,font=\sffamily\Large\bfseries}]

  \node[main node] (As) {As};
  \node[main node] (Ni) [above right of=As] {Ni};
  \node[main node] (Fe) [below right of=As] {Fe};
  \node[main node] (Cr) [below right of=Ni] {Cr};
  \node[main node] (Zn) [right of=Cr] {Zn};

  \path[every node/.style={font=\sffamily\small}]
    (Zn) edge node [above]{+} (Cr)
    (Fe) edge node [right]{+} (Cr)
        edge node [below] {+} (As)
        edge node [right] {+} (Ni)
    (Cr) edge node [right] {+} (Ni)
    (Ni) edge node [above] {-} (As);
\end{tikzpicture}
\caption{Reduced conditional dependence graph for the five soil components.}
\label{fig:graph}
\end{figure}
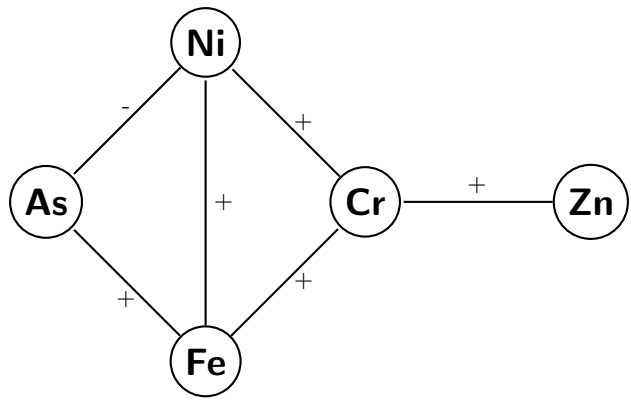

To illustrate these conditional relationships more intuitively, a conditional dependence graph is provided in Figure \ref{fig:graph}. Each of the soil elements is represented by a node, with edges connecting the nodes when there is a positive (indicated by a `+') or negative (indicated by a `-') conditional relationship between the elements. For example, there is no edge connecting arsenic and chromium, suggesting that they are independent conditional on nickel and iron. That is, the observed pairwise correlation between arsenic and chromium can be explained statistically via the relationships each have with nickel and iron. 

Finally, as a form of model-checking, we can compare the analysis results for arsenic, iron and chromium to those of \cite{guinness2014multivariate}. The covariance functions in Figure \ref{fig:covar} are consistent with those found in the previous work, as are the shapes of the cross-covariance functions in Figure \ref{fig:crosscovar}. This agreement is encouraging and suggests that minimal meaningful flexibility was lost by setting $\boldsymbol\rho(\om)\equiv \boldsymbol\rho$. Additionally, \cite{guinness2014multivariate} explored the dependence between arsenic and chromium through a hypothesis testing approach and found ``strong but not overwhelming evidence for conditional dependence." In contrast, our analysis suggests that when nickel and zinc are incorporated into the model, the conditional dependence between chromium and arsenic is no longer significant. I.e., not surprisingly, the nature of these conditional dependencies will change depending on the soil elements being accounted for. This highlights the importance of being able to accommodate several variables of interest, previously impossible using the earlier model.

Our model can also be checked with respect to the known chemistry of arsenic. Scanning electron microscopy - energy-dispersive X-ray (SEM-EDX) analysis indicated that visibly dark spots on the sand-grain sample corresponded with hotspots (Figure \ref{fig:data}) that are enriched in iron, chromium, and nickel at a ratio corresponding with stainless steel (SEM-EDX data not shown). We deduced that the contamination was introduced to the sample during mounting with stainless steel forceps. Stainless steel adsorbs arsenate, which explains the suppressed arsenic accumulation at the locations of iron-chromium-nickel co-enrichment. Since the largest relative abundance of nickel was due to the stainless steel, this contamination explains the negative conditional relationship observed between arsenic and nickel. In contrast, the marginal and conditional relationships between arsenic and iron were both identified to be positive. This result can be explained by noting that the most common form of iron in the sample (Fe(III)) was natural and has a high capacity to bind arsenic, and that the competing effects of arsenic adsorption by stainless steel is known to be trivial in comparison.
%However, arsenic adsorption on stainless steel should be trivial compared with that of soil iron oxides. It is not surprising then that despite the presence of iron in the stainless steel, the marginal and conditional relationships between arsenic and iron were both identified to be positive. This result can more specifically be explained by noting that the most common form of iron in the sample (Fe(III)) was natural and has a high capacity to bind arsenic, and that the arsenic adsorption on stainless steel is known to be trivial in comparison. % % %%Considering Fe, Cr, and Ni fluorescence data across the entire ROI modeled, which included native elements and stainless steel, the negative relationship of As-Fe conditioned on Cr \cite[]{guinness2014multivariate} or on Ni (this work) makes chemical sense with respect to this contamination effect.

% % % % % % % % % % % % % % % % % % % %
\section{Discussion}
\label{sec:concl}
The framework we have outlined enables joint spatial modeling of multiple spatial processes in a computationally efficient manner through the use of Fourier transformations and spectral analysis theory. The non-separable covariance structure allows a unique covariance function for each of the variables, and a unique cross-covariance function for each pair of variables. By constructing the model in a Bayesian setup, we can fully quantify the uncertainty associated with our estimates and make inference on the strength of pairwise variable relationships. Post-model fitting examination of the correlation matrix additionally provides inference on the conditional relationships between the variables.

%This approach was illustrated for five soil elements measured on a sand grain using $\mu$XRF analysis. Understanding interactions between these soil elements has the potential to improve the efficacy of relief efforts for widespread arsenic poisoning in Asia and around the world, but these relationships have proven difficult to study in laboratory settings. Here we are able to provide statistical inference on the pairwise relationships between various soil elements, and can go one step further by providing a full description of all conditional relationships. Previous work analyzing these data was unable to account for more than three soil elements at a time and lacked adequate descriptions of uncertainty \cite[]{guinness2014multivariate}, neither of which is the case under this new approach.

This approach was illustrated for soil elements mapped in thin mineral-organic coatings on the surface of a quartz soil-sand grain using $\mu$-XRF analysis, including elements that are known to occur at least partially as non-reactive stainless-steel contamination. Here we are able to provide statistical inference on the pairwise relationships between five soil elements, and can go one step further by providing a full description of all conditional relationships. Unlike our new approach, previous work analyzing these data was not able to account for more than three soil elements at a time and also lacked adequate descriptions of uncertainty \cite[]{guinness2014multivariate}.

Our modeling approach provides a framework for assessing conditioning of multiple soil elements that can have different geochemical effects on arsenic accumulation, and analyses here could reasonably be expanded with additional spatial data sets for elements such as aluminum, manganese, and carbon, which also are known to impact arsenic reactivity. Ultimately, the approach developed here could provide a useful tool for more broadly regulating negative impacts of environmental pollutants. Our current understanding of mechanisms of chemical-contaminant reactivity is largely gleaned from laboratory experiments involving pure, simplified systems. Interactive effects between pollutants, multiple minerals and/or organic matter co-localized in soil microsites are difficult to disentangle using current analytical techniques, thereby limiting our predictive capabilities in natural environmental settings. An understanding of these interactions from experiments conducted directly with soil materials would allow us to formulate hypotheses of chemical reactivity in complex systems and improve mechanistic models that, for instance, help predict the soil and sediment controls on arsenic poisoning of Asian well water. 

In its current form the proposed methodology is appropriate for jointly modeling multiple spatial variables observed on a complete lattice. However, future adaptations can be envisioned to accommodate such spatial processes that are non-stationary \cite[]{fuentes2002spectral}, and/or were collected on an incomplete lattice \cite[]{fuentes2007approximate, stroud2014bayesian}. 

% % % % % % % % % % % % % % % % %
% % % % % % % % % %
\bibliographystyle{ba}
\bibliography{SoilModel}

%\begin{acknowledgement}
%The authors thank Joseph Guinness for helpful discussions regarding the spectral analysis theory. Terres and Fuentes are supported by the National Science Foundation's Research Network for Statistical Methods for Atmospheric and Oceanic Sciences, award number 1107046.
%\end{acknowledgement}

\end{document}